\documentclass[12pt,epsfig]{article}
\newcommand{\ba}{\begin{array}{c}}
\newcommand{\baz}{\begin{array}{cc}}
\newcommand{\bad}{\begin{array}{ccc}}
\newcommand{\bav}{\begin{array}{cccc}}
\newcommand{\ea}{\end{array}}

\usepackage{amssymb} 
\usepackage{epsfig}
\usepackage{float}
\newcommand{\be}{\begin{equation}}
\newcommand{\ee}{\end{equation}}
\newcommand{\bea}{\begin{eqnarray}}
\newcommand{\eea}{\end{eqnarray}}
\begin{document}

\begin{center}
\bf {The quark mixing matrix with manifest Cabibbo substructure
and an angle of the unitarity triangle as one of its parameters}
\end{center}

\begin{center}
C. Jarlskog 
\end{center}

\begin{center}
{\em Division of Mathematical Physics\\
LTH, Lund University\\
Box 118, S-22100 Lund, Sweden}
\end{center}

\begin{abstract}

The quark mixing matrix is parameterised such that its "Cabibbo substructure"
is emphasised. One can choose one of the parameters to be an arbitrarily
chosen angle of 
the unitarity triangle, for example the angle $\beta$
(also called $\Phi_1$).  

\end{abstract}

\section{Introduction}

The question of fermion masses and mixings has been among the
most central issues in theoretical particle physics since a long time. 
Within the three family version of the Standard Model \cite{sm} many
specific forms for the quark mass matrices have been proposed in the past
with the hope that some insight may be gained into the flavour problem.
For example, already in 1978 Fritzsch \cite{fritzsch} proposed a structure 
which became quite popular as it could be realised in some Grand Unified
Theories (see, for example Ref. \cite{georgi}).
Since then possible zeros in the quark mass matrices (usually called texture zeros) 
have enjoyed special popularity as these make the computations
more transparent and generally lead 
to specific predictions. Again one has hoped that clues to the solution
of the flavour problem may emerge. Another approach
has been to "derive" quark mass matrices
from experiments, see, for example Ref. \cite{frampton} 
where it was found that the two quark mass
matrices are highly "aligned".
  
A troubling factor in all such studies is that the mass matrices are not uniquely 
defined but are "frame" dependent.
In other words, given any set of three-by-three quark mass matrices
$M_u$ and $M_d$, 
for the up-type and down-type quarks respectively, one can obtain other sets by
unitary rotations without affecting the physics. The measurables are, of course,
frame-independent and therefore they must be invariant functions under such 
unitary rotations. 
These functions were introduced in \cite{ceja85a} and studied in 
detail in \cite{ceja87}. Furthermore, it has been shown recently \cite{ceja05} 
that this formalism can be extended to
the case of neutrino oscillations.
 
For the quarks what enters, in the standard model, is the pair
\begin{equation} 
S_u \equiv M_u M_u^\dagger~,~S_d \equiv M_d M_d^\dagger 
\end{equation}

The original motivation for the work presented here was to 
look for "the golden mean" mass matrices, to be defined shortly.
First we note that there are two "extreme frames", 
one in which the up-type quark mass matrix is diagonal, i.e., 
\begin{equation} 
S_u =
\left( \begin{array}{ccc}
m^2_u & 0 & 0 \\
0 & m^2_c & 0 \\
0 & 0 & m^2_t
\end{array}
\right), ~~~~~ S_d = V
\left( \begin{array}{ccc}
m^2_d & 0 & 0 \\
0 & m^2_s & 0 \\
0 & 0 & m^2_b
\end{array}
\right) V^\dagger 
\end{equation}
where the $m$'s refer to the quark masses and $V$ is the quark mixing matrix. 
The other extreme frame is one in which the down-type quark mass matrix is 
diagonal, i.e.,
\begin{equation}
S_d = 
\left( \begin{array}{ccc}
m^2_d & 0 & 0 \\
0 & m^2_s & 0 \\
0 & 0 & m^2_b
\end{array}
\right), ~~~~~   
S_u = V^\dagger
\left( \begin{array}{ccc}
m^2_u & 0 & 0 \\
0 & m^2_c & 0 \\
0 & 0 & m^2_t
\end{array}
\right) V,  
\end{equation}
One may then wonder how the mass matrices would look like in the
"golden mean frame", i.e., the frame right in the middle of the two extremes, where
\begin{equation} 
S_u = W^\dagger
\left( \begin{array}{ccc}
m^2_u & 0 & 0 \\
0 & m^2_c & 0 \\
0 & 0 & m^2_t
\end{array}
\right) W, ~~~~~ S_d = W
\left( \begin{array}{ccc}
m^2_d & 0 & 0 \\
0 & m^2_s & 0 \\
0 & 0 & m^2_b
\end{array}
\right) W^\dagger 
\end{equation}
$W$ is the square root of the quark mixing matrix,
\begin{equation}
V = W^2
\end{equation}
In order to go to this frame one needs to compute the square root of the
quark mixing matrix. The specific parameterisation of 
$V$ turns out to be of paramount importance for achieving this goal.
In spite of the fact that all valid parameterisations are 
physically equivalent, most of them are "nasty" and don't allow 
their roots to be taken
so easily. After several attempts and having got stopped by heavy calculations, 
we have found a particularly convenient parameterisation,
presented here below. It turns out that this parameterisation by itself 
is more interesting than the answer to our original question, which will
be dealt with in a future publication.

\section{A parameterisation with manifest Cabibbo substructure}

The quark mixing matrix is usually parameterised as a function of three rotation
angles and one phase, generally denoted by the set  
$\theta_1, \theta_2, \theta_3$ and $\delta$. 
However there are many ways in which these parameters can be introduced (for a
review see, for example \cite{cpboken}) and the meaning of these quantities
depends on how they are introduced.
A specific parameterisation may have some beautiful features as well as short-comings.
For example, a special feature of the seminal 
Kobayashi-Maskawa parameterisation \cite{km} 
is that in the limit $\theta_1 \rightarrow 0$ 
the first family decouples from the other two.
The parameterisation preferred by the Particle Data Group \cite{pdg} has as its 
special feature that its phase $\delta$ is locked to the smallest angle $\theta_3$
but none of the families decouples if only one of the angles goes to zero.
A most important and easy to remember empirical parameterisation 
has been given by Wolfenstein \cite{lincoln}, 
where the matrix is expanded in
powers of a parameter denoted by $\lambda$, where $\lambda \simeq 0.22$. 

In this article, we introduce an (exact) parameterisation of 
the quark mixing matrix in terms of four parameters denoted by
$\Phi, \theta_3, \delta_\alpha$ and $\delta_\beta$. The reason for calling one
of the angles $\theta_3$ when we have no other $\theta$'s is to
stay as close as possible to the usual nomenclature. Our angles
$\delta$ are often somewhat different from what is 
commonly used and thus, in
order not to confuse the reader, we do not denote them with $\theta$.

We write the quark mixing matrix (exactly) in a form such that
its Cabibbo substructure is emphasised from the very beginning,
\begin{equation} V = V_0 + s_3 V_1 + (1-c_3) V_2
\label{defV} 
\end{equation}
where $s_3 = sin\theta_3$, $c_3 = cos \theta_3$ and the matrices 
$V_j$, $j=0-2$, are given by
\begin{eqnarray} 
V_0 &=&
\left( \begin{array}{ccc}
cos \Phi & sin \Phi & 0 \\
-sin \Phi & cos \Phi & 0 \\
0 & 0 & 1
\end{array}
\right) = \left( \begin{array}{cc} R_2(\Phi) & 
\begin{array}{c} 0 \\ 0 \end{array} \\ \begin{array}{cc}
0~& 0 \end{array} & 1 
\end{array}
\right)  \\
 V_1 &=&
\left( \begin{array}{ccc}
0& 0 & a_1 \\
0 & 0 & a_2\\
b_1^\star & b_2^\star & 0
\end{array}
\right) \equiv \left( \begin{array}{cc} 0 & \vert A> \\ <B \vert & 0 
\end{array} \right)\\
 V_2 &=&
\left( \begin{array}{cc} \vert A > < B \vert & 
\begin{array}{c} 0 \\ 0 \end{array} \\ \begin{array}{cc}
0~& 0 \end{array} & -1 
\end{array}
\right) 
\end{eqnarray}
Here
\begin{equation}
\vert A >= \left( \begin{array}{c} a_1 \\ a_2 \end{array} \right) ~~~~
\vert B >= \left( \begin{array}{c} b_1 \\ b_2 \end{array} \right)
\end{equation}
and $(\vert A > < B \vert)_{ij} \equiv a_i b_j^\star$. 
We will impose the following conditions on $A$ and $B$: 
\begin{equation}
<A \vert A> = <B \vert B> =1
\end{equation}
and
\begin{eqnarray}
\vert A> &=& - R_2 (\Phi) \vert B> \nonumber \\
\vert B> &=& - R_2 (-\Phi) \vert A>
\label{relab}
\end{eqnarray}
By these conditions, the vector $A$ represents two real parameters,
for example the magnitude of $a_1$ and the relative 
phase of $a_1$ and $a_2$. These will provide the two remaining
parameters ($\delta_\alpha, \delta_\beta$) that together
with $\Phi$ and $\theta_3$ add up to the four parameters
needed to get the most general quark mixing matrix. 
Because of Eq. (\ref{relab}) $B$ 
introduces no further parameters. 
Note that 
\begin{equation}
V_{13} = a_1 s_3, ~~V_{23} = a_2 s_3, ~~
V_{31} = b^\star_1 s_3, ~~
V_{32} = b^\star_2 s_3
\end{equation}

We will also introduce the invariant $J$ defined by
\begin{equation}
Im (V_{\alpha j}V_{\beta k}V^\star_{\alpha k}V^\star_{\beta j})= J~\sum_{\gamma , l}^{}
\epsilon_{\alpha\beta\gamma} \epsilon_{jkl}
\label{defj}
\end{equation}

In the above parameterisation we find
\begin{equation}
J = s^2_3 c_3 sin\Phi cos\Phi Im(a_1^\star a_2) = 
s^2_3 c_3 sin\Phi cos\Phi Im(b_1^\star b_2)
\end{equation}
where the last equality follows from Eq.(\ref{relab}).

We can check the unitarity of the matrix $V$ without specifying
what $A$ (or equivalently $B$) looks like. We find

\[V_0 V_1^\dagger + V_1 V_0^\dagger = V_1 V_2^\dagger + V_2 V_1^\dagger =0  \]

 \[ V_2 V_2^\dagger = V_1 V_1^\dagger = -{1 \over 2}(V_0 V_2^\dagger + V_2 V_0^\dagger) =  
\left( \begin{array}{cc} \vert A > < A \vert & 
\begin{array}{c} 0 \\ 0 \end{array} \\ \begin{array}{cc}
0~& 0 \end{array} & 1 
\end{array}
\right)
\]
These identities are derived trivially by using the relation between 
$A$ and $B$, Eq.(\ref{relab}).

Given any $A$ or $B$ we have the freedom to rephase it,
for example
\begin{equation}
\vert A> \rightarrow e^{i\eta} \vert A>
\end{equation}
whereby the vector $B$ is also rephased by the same amount
(see Eq.(\ref{relab})). From the form of the matrix $V$ we
see immediately that the elements $V_{11}, V_{12}, V_{21}, V_{22}$ 
and $V_{33}$ remain invariant under this rephasing. 

In this parameterisation, the usual unitarity triangle, obtained
from Eq.(\ref{relab}), is a consequence of
\begin{equation}
a_1 cos\Phi - a_2 sin\Phi + b_1 =0
\end{equation}
Thus the three angles of the triangle are given by the 
phases of $b_1 a_2^\star$,
$a_2 a_1^\star$ and $a_1 b_1^\star$. 
We can choose our $A$ or $B$ such that one of these angles enters
directly as a parameter in the matrix $V$. The simplest one to 
incorporate is the angle usually denoted by $\gamma$ i.e., the
phase of $a_2 a_1^\star$. We could choose
\begin{equation}
\vert A > = \left( \begin{array}{c}
sin\delta_\beta e^{-i \delta_\alpha}\\ cos \delta_\beta
\end{array} \right) 
\end{equation}
whereby
\[sin\delta_\alpha = sin\gamma,~~~
J = s^2_3 c_3 sin\Phi cos\Phi sin\delta_\beta cos \delta_\beta
sin\gamma  \]
We would then compute $B$ using Eq.(\ref{relab}).

To incorporate the angle $\beta$ (also denoted by $\phi_1$) 
of the unitarity triangle
we could take $a_2$ to be real and $b_1$
to have the phase $\delta_\beta = \beta$. 
From Eq.(\ref{relab}), the reality 
condition on $a_2$ implies that $-sin\Phi b_1 + cos\Phi b_2$ be real.
This fixes the vector $B$ and thereby also the vector $A$. We find
\begin{equation} 
\vert B>  = {1 \over \sigma}
\left( \begin{array}{c} cos\Phi sin\delta_\alpha e^{i\delta_\beta} \\ 
-sin\Phi sin\delta_\beta e^{-i\delta_\alpha} \end{array} \right)
\label{b} 
\end{equation}
where
\begin{equation} 
\sigma^2 = cos^2\Phi sin^2\delta_\alpha + sin^2\Phi sin^2\delta_\beta
\end{equation}
The vector $A$ thus obtained is given by
 \begin{equation} 
\vert A> = {1 \over 2\sigma } \left( \begin{array}{c} 
 -[cos2\Phi sin(\delta_\alpha +\delta_\beta )
+ sin\delta_\alpha e^{i\delta_\beta} - sin\delta_\beta e^{-i\delta_\alpha}]
 \\ sin 2\Phi sin(\delta_\alpha +\delta_\beta ) 
\end{array} \right)
\label{a}
\end{equation} 
Here   
\begin{equation}
sin \delta_\beta = sin \beta (BABAR) = sin\Phi_1 (BELLE)
\end{equation}
where BABAR \cite{babar} and BELLE \cite{belle} Collaborations have 
determined this angle
in their study of the $B-\bar{B}$ system but use different notations
for it.

With this choice, $J$ is given by
\begin{equation} 
J = s_3^2 c_3 {sin^2(2\Phi)sin\delta_\alpha
sin\delta_\beta sin(\delta_\alpha +
\delta_\beta) \over 4 \sigma^2}
\end{equation}

Finally in order to utilise the third angle, $\alpha$ also
known as $\phi_2$, as a parameter we may take it to be the phase of $b_1$
and require that $a_1$ be real. The procedure to be followed
to achieve this goal is exactly as depicted above. 

The above expressions may look somewhat complicated but they are 
generally quite easy to work with as we often
only need their closed forms and not their details. 
 
\section{Special features and an estimation of the parameters}
 
The above parameterisation, Eq.(\ref{defV}), is an {\it exact} form and
not a perturbative expansion. It has several special features as follows:

1. In the limit $\theta_3 \rightarrow 0$ the
third family decouples from the first two and the exact 
Cabibbo substructure, with the mixing angle $\Phi$ 
between the first two families, emerges 

2. Since the matrices $V_j, j=0-2$,
{\it do not } depend on $\theta_3$, this parameterisation provides
a convenient framework for perturbative expansion 
in powers of $\theta_3$ which is indeed small, 
of order $\lambda^2$. 

3. We have seen that we can incorporate any one of the angles of
the unitarity triangle as one of the four parameters of the mixing matrix.

We now estimate the value of our parameters $\Phi, \theta_3, 
\delta_\alpha, \delta_\beta$ for the choice
Eq.(\ref{b}) by comparing them with Wolfenstein's parameters 
 \cite{lincoln}. Comparing the matrix elements
$V_{12}$ and $V_{33}$ yields that the angles $\Phi$ and $\theta_3$
are or order $\lambda$ and $\lambda^2$ respectively
\begin{equation}
\Phi \simeq \lambda, ~~~ \theta_3 \simeq A \lambda^2 
\end{equation}
Next, from the moduli of the matrix elements 
$V_{13}$, $V_{23}$, $V_{31}$, and $V_{32}$ we find
that the angle $\delta_\alpha$ is much smaller than the angle $\delta_\beta$,
\begin{eqnarray}
sin\delta_\beta&\simeq{\eta \over \sqrt{(1-\rho)^2 + \eta^2}}\\
cos\delta_\beta&\simeq{1-\rho \over \sqrt{(1-\rho)^2 + \eta^2}} \\
\end{eqnarray}
\begin{equation}
sin\delta_\alpha \simeq \eta \lambda^2
\end{equation}
Finally, the invariant $J$ is given by
\begin{equation}
J \simeq \theta_3^2 sin\delta_\alpha = A^2 \lambda^4 sin\delta_\alpha
\end{equation}

There is a somewhat subtle issue about this parameterisation that merits to be
discussed even though it is hypothetical. It concerns
the case with CP conservation while we know that CP is violated 
and therefore the parameters 
$\delta_\alpha$ and $\delta_\beta$ are both nonvanishing. Nonetheless,
we are used to parameterisations with
three rotation angles and a phase such that when the phase approaches
zero one immediately obtains a mixing matrix with three rotation angles.
The converse is not necessarily true that when one of the angles vanishes
so does the phase. To remove the phase one often needs to expend some effort.
The parameterisation here is more like having two rotation angles and
two phases; both of the latter vanish when there is no CP violation. It
would seem that we would end up with only two angles, $\Phi$ and
$\theta_3$. 
 How do we then recover the third angle, which should be there?

The answer is that even though in the CP conserving limit 
$\delta_\alpha$ and $\delta_\beta$ both approach zero 
their ratio needs to be defined. 
We may introduce two angles,
$\theta_1$ and $\theta_2$, by putting
\begin{eqnarray}
\Phi &=& \theta_1 + \theta_2 \\
{sin \delta_\alpha \over sin\delta_\beta} &=& 
tan \theta_1 ~tan(\theta_1+\theta_2) 
\end{eqnarray}
Taking the limits carefully as the two $\delta$'s approach zero, we find
\begin{equation}
\vert B> = \left( \begin{array}{c} sin\theta_1 \\ -cos\theta_1
\end{array} \right),~~~
\vert A> = \left( \begin{array}{c} sin\theta_2 \\ cos\theta_2
\end{array} \right)
\end{equation}
and thus we end up with a mixing matrix with just three rotation
angles. Furthermore, in this limit the invariant $J$ contains three powers
of $sin\delta$ ($\delta$ being $\delta_\alpha$ or $\delta_\beta$)
in its numerator but only two 
in its denominator and thus vanishes as it should.

\end{document}